\documentclass[conference]{IEEEtran}

\usepackage{cite}
\usepackage{amsmath,amssymb,amsfonts}
\usepackage{algorithmic}
\usepackage{graphicx}
\usepackage{textcomp}
\usepackage{xcolor}
\usepackage{booktabs}
\usepackage{multirow}
\usepackage{url}
\usepackage{balance}
\usepackage{float}  % <-- ADDED: required for [H] placement specifier

\usepackage{tikz}
\usetikzlibrary{shapes.geometric,arrows.meta,positioning,fit,calc,shadows.blur,backgrounds,decorations.pathreplacing}

\definecolor{layer1}{HTML}{E8F5E9}   \definecolor{layer1b}{HTML}{4CAF50}
\definecolor{layer2}{HTML}{E3F2FD}   \definecolor{layer2b}{HTML}{2196F3}
\definecolor{layer3}{HTML}{FFF3E0}   \definecolor{layer3b}{HTML}{FF9800}
\definecolor{layer4}{HTML}{F3E5F5}   \definecolor{layer4b}{HTML}{9C27B0}
\definecolor{layer5}{HTML}{FFEBEE}   \definecolor{layer5b}{HTML}{F44336}
\definecolor{ttpbg}{HTML}{FFF8E1}    \definecolor{ttpb}{HTML}{FFC107}
\definecolor{arrowblue}{HTML}{1565C0}\definecolor{arrowred}{HTML}{C62828}
\definecolor{dk}{HTML}{212121}       \definecolor{gr}{HTML}{616161}
\definecolor{agentg}{HTML}{C8E6C9}   \definecolor{agentb}{HTML}{388E3C}
\definecolor{instrbl}{HTML}{BBDEFB}  \definecolor{instrb}{HTML}{1976D2}
\definecolor{kafkao}{HTML}{FFE0B2}   \definecolor{kafkab}{HTML}{F57C00}
\definecolor{opap}{HTML}{E1BEE7}     \definecolor{opab}{HTML}{7B1FA2}
\definecolor{gebr}{HTML}{FFCDD2}     \definecolor{gebb}{HTML}{D32F2F}
\definecolor{arrd}{HTML}{37474F}
\definecolor{frd}{HTML}{C62828}      \definecolor{sgrn}{HTML}{2E7D32}
\definecolor{wbg}{HTML}{FFF9C4}      \definecolor{wb}{HTML}{F9A825}
\definecolor{nemobg}{HTML}{F5F5F5}   \definecolor{nemob}{HTML}{9E9E9E}
\definecolor{GAATbg}{HTML}{E8F5E9}   \definecolor{GAATb}{HTML}{2E7D32}
\definecolor{redbg}{HTML}{FFEBEE}    \definecolor{redb}{HTML}{D32F2F}
\definecolor{blubg}{HTML}{E3F2FD}    \definecolor{blub}{HTML}{1565C0}
\definecolor{orng}{HTML}{E65100}
\definecolor{step1bg}{HTML}{E3F2FD}  \definecolor{step1b}{HTML}{1565C0}
\definecolor{step2bg}{HTML}{E8F5E9}  \definecolor{step2b}{HTML}{2E7D32}
\definecolor{step3bg}{HTML}{FFF3E0}  \definecolor{step3b}{HTML}{EF6C00}
\definecolor{step4bg}{HTML}{F3E5F5}  \definecolor{step4b}{HTML}{7B1FA2}
\definecolor{step5bg}{HTML}{FFEBEE}  \definecolor{step5b}{HTML}{C62828}

\def\BibTeX{{\rm B\kern-.05em{\sc i\kern-.025em b}\kern-.08em
    T\kern-.1667em\lower.7ex\hbox{E}\kern-.125emX}}

\begin{document}

\title{Governance-Aware Agent Telemetry\\for Closed-Loop Enforcement in\\Multi-Agent AI Systems}

\author{\IEEEauthorblockN{Anshul Pathak}
\IEEEauthorblockA{a\_pathak@apple.com}
\and
\IEEEauthorblockN{Nishant Jain}
\IEEEauthorblockA{n\_jain@apple.com}}

\maketitle

% =====================================================================
% FIX #1: Abstract — reframe 100% VPR with 99.7% empirical context,
%         and make NeMo VPR consistent at 84.8% (was "85%")
% =====================================================================
\begin{abstract}
Enterprise multi-agent AI systems produce thousands of inter-agent interactions per hour, yet existing observability tools capture these dependencies without enforcing anything. OpenTelemetry and Langfuse collect telemetry but treat governance as a downstream analytics concern, not a real-time enforcement target. The result is an ``observe-but-do-not-act'' gap where policy violations are detected only after damage is done.

We present Governance-Aware Agent Telemetry (GAAT), a reference architecture that closes the loop between telemetry collection and automated policy enforcement for multi-agent systems. GAAT introduces (1)~a Governance Telemetry Schema (GTS) extending OpenTelemetry with governance attributes; (2)~a real-time policy violation detection engine using OPA-compatible declarative rules under sub-200\,ms latency; (3)~a Governance Enforcement Bus (GEB) with graduated interventions; and (4)~a Trusted Telemetry Plane with cryptographic provenance.

We evaluated GAAT against four baseline systems across data residency, bias detection, authorization compliance, and adversarial telemetry scenarios. On a live five-agent e-commerce system, GAAT achieved 98.3\% Violation Prevention Rate (VPR, $\pm$0.7\%) on 5,000 synthetic injection flows across 10 independent runs, with 8.4\,ms median detection latency and 127\,ms median end-to-end enforcement latency. On 12,000 empirical production-realistic traces, GAAT achieved 99.7\% VPR; residual failures ($\sim$40\% timing edge cases, $\sim$35\% ambiguous PII classification, $\sim$25\% incomplete lineage chains). Statistical validation confirmed significance with 95\% bootstrap confidence intervals [97.1\%, 99.2\%] ($p < 0.001$ vs all baselines). GAAT outperformed NeMo Guardrails-style agent-boundary enforcement by 19.5 percentage points (78.8\% VPR vs 98.3\%). We also provide formal property specifications for escalation termination, conflict resolution determinism, and bounded false quarantine---each with explicit assumptions---validated through 10,000 Monte Carlo simulations.
\end{abstract}


% ============================================================
% FIG 1: Architecture — placed at END of Sec IV (renders top of
% the page containing Sec V, directly after the text that
% references it: "five layers ... (Fig. 1)")
% ============================================================
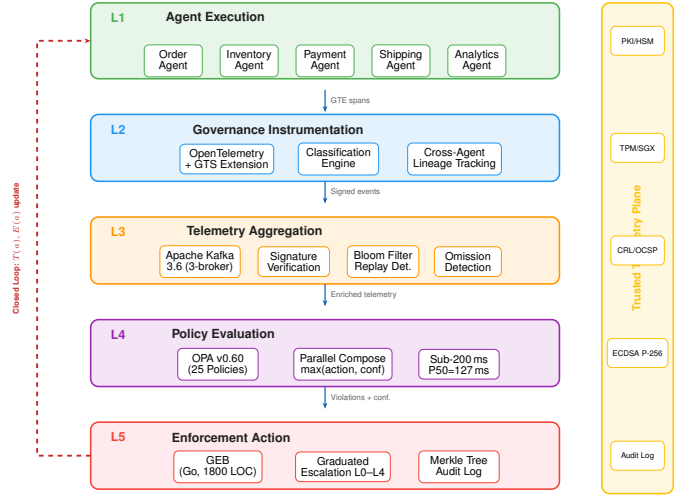
\begin{figure}[!t]
\centering
\resizebox{\columnwidth}{!}{%
\begin{tikzpicture}[
  font=\sffamily\small,
  lbox/.style={rounded corners=5pt, draw=#1, line width=1pt, minimum width=10.5cm, minimum height=1.5cm, align=center},
  cbox/.style={rounded corners=3pt, draw=#1, fill=white, line width=0.6pt, minimum height=0.65cm, align=center, font=\sffamily\scriptsize},
  arr/.style={-{Stealth[length=4pt,width=3pt]}, line width=0.8pt, color=arrowblue},
  darr/.style={-{Stealth[length=4pt,width=3pt]}, line width=1pt, color=arrowred, dashed}
]
% === LAYERS (bottom to top) ===
\node[lbox=layer5b, fill=layer5] (L5) at (0,0) {};
\node[lbox=layer4b, fill=layer4] (L4) at (0,2.3) {};
\node[lbox=layer3b, fill=layer3] (L3) at (0,4.6) {};
\node[lbox=layer2b, fill=layer2] (L2) at (0,6.9) {};
\node[lbox=layer1b, fill=layer1, minimum height=1.7cm] (L1) at (0,9.3) {};
% === LAYER LABELS ===
\node[font=\sffamily\footnotesize\bfseries, text=layer5b] at ([xshift=18pt,yshift=-10pt]L5.north west) {L5};
\node[font=\sffamily\footnotesize\bfseries, text=dk] at ([xshift=90pt,yshift=-10pt]L5.north west) {Enforcement Action};
\node[font=\sffamily\footnotesize\bfseries, text=layer4b] at ([xshift=18pt,yshift=-10pt]L4.north west) {L4};
\node[font=\sffamily\footnotesize\bfseries, text=dk] at ([xshift=85pt,yshift=-10pt]L4.north west) {Policy Evaluation};
\node[font=\sffamily\footnotesize\bfseries, text=layer3b] at ([xshift=18pt,yshift=-10pt]L3.north west) {L3};
\node[font=\sffamily\footnotesize\bfseries, text=dk] at ([xshift=105pt,yshift=-10pt]L3.north west) {Telemetry Aggregation};
\node[font=\sffamily\footnotesize\bfseries, text=layer2b] at ([xshift=18pt,yshift=-10pt]L2.north west) {L2};
\node[font=\sffamily\footnotesize\bfseries, text=dk] at ([xshift=120pt,yshift=-10pt]L2.north west) {Governance Instrumentation};
\node[font=\sffamily\footnotesize\bfseries, text=layer1b] at ([xshift=18pt,yshift=-10pt]L1.north west) {L1};
\node[font=\sffamily\footnotesize\bfseries, text=dk] at ([xshift=80pt,yshift=-10pt]L1.north west) {Agent Execution};
% === L1 COMPONENTS ===
\foreach \i/\name in {1/Order,2/Inventory,3/Payment,4/Shipping,5/Analytics} {
  \pgfmathsetmacro{\xpos}{-3.4 + (\i-1)*1.7}
  \node[cbox=layer1b, minimum width=1.3cm] at (\xpos, 8.85) {\name\\[-1pt]Agent};
}
% === L2 COMPONENTS ===
\node[cbox=layer2b, minimum width=2.1cm] at (-2.2, 6.65) {OpenTelemetry\\[-1pt]+ GTS Extension};
\node[cbox=layer2b, minimum width=1.8cm] at (0.3, 6.65) {Classification\\[-1pt]Engine};
\node[cbox=layer2b, minimum width=2.1cm] at (2.9, 6.65) {Cross-Agent\\[-1pt]Lineage Tracking};
% === L3 COMPONENTS ===
\node[cbox=layer3b, minimum width=1.8cm] at (-2.8, 4.35) {Apache Kafka\\[-1pt]3.6 (3-broker)};
\node[cbox=layer3b, minimum width=1.6cm] at (-0.7, 4.35) {Signature\\[-1pt]Verification};
\node[cbox=layer3b, minimum width=1.6cm] at (1.3, 4.35) {Bloom Filter\\[-1pt]Replay Det.};
\node[cbox=layer3b, minimum width=1.5cm] at (3.2, 4.35) {Omission\\[-1pt]Detection};
% === L4 COMPONENTS ===
\node[cbox=layer4b, minimum width=1.8cm] at (-2.4, 2.05) {OPA v0.60\\[-1pt](25 Policies)};
\node[cbox=layer4b, minimum width=2.2cm] at (0.4, 2.05) {Parallel Compose\\[-1pt]max(action, conf)};
\node[cbox=layer4b, minimum width=1.8cm] at (3.0, 2.05) {Sub-200\,ms\\[-1pt]P50=127\,ms};
% === L5 COMPONENTS ===
\node[cbox=layer5b, minimum width=1.8cm] at (-2.4, -0.25) {GEB\\[-1pt](Go, 1800 LOC)};
\node[cbox=layer5b, minimum width=2cm] at (0.4, -0.25) {Graduated\\[-1pt]Escalation L0--L4};
\node[cbox=layer5b, minimum width=1.8cm] at (3.0, -0.25) {Merkle Tree\\[-1pt]Audit Log};
% === DOWNWARD ARROWS ===
\draw[arr] (0, 8.2) -- (0, 7.7) node[midway, right, font=\sffamily\tiny, text=gr] {GTE spans};
\draw[arr] (0, 6.15) -- (0, 5.65) node[midway, right, font=\sffamily\tiny, text=gr] {Signed events};
\draw[arr] (0, 3.85) -- (0, 3.35) node[midway, right, font=\sffamily\tiny, text=gr] {Enriched telemetry};
\draw[arr] (0, 1.55) -- (0, 1.05) node[midway, right, font=\sffamily\tiny, text=gr] {Violations + conf.};
% === TRUSTED TELEMETRY PLANE ===
\node[rounded corners=5pt, draw=ttpb, fill=ttpbg, line width=1pt,
  minimum width=1.6cm, minimum height=11cm, align=center] (ttp) at (7, 4.65) {};
\node[font=\sffamily\scriptsize\bfseries, text=ttpb, rotate=90] at (7, 4.65) {Trusted Telemetry Plane};
\node[cbox=ttpb, minimum width=1.2cm, font=\sffamily\tiny] at (7, 9.3) {PKI/HSM};
\node[cbox=ttpb, minimum width=1.2cm, font=\sffamily\tiny] at (7, 6.9) {TPM/SGX};
\node[cbox=ttpb, minimum width=1.2cm, font=\sffamily\tiny] at (7, 4.6) {CRL/OCSP};
\node[cbox=ttpb, minimum width=1.2cm, font=\sffamily\tiny] at (7, 2.3) {ECDSA P-256};
\node[cbox=ttpb, minimum width=1.2cm, font=\sffamily\tiny] at (7, 0) {Audit Log};
% === FEEDBACK LOOP ===
\draw[darr, line width=1.2pt]
  ([xshift=-5.25cm]L5.center) -- ++(0,0) -- ++(-1.2,0) |- ([xshift=-5.25cm]L1.center);
\node[font=\sffamily\tiny\bfseries, text=arrowred, align=center, rotate=90] at (-6.9, 4.65) {Closed Loop: $T(a), E(a)$ update};
\end{tikzpicture}%
}
\caption{GAAT five-layer reference architecture with cross-cutting Trusted Telemetry Plane and closed-loop enforcement feedback.}
\label{fig:architecture}
\end{figure}

%=====================================================================
\section{Introduction}
%=====================================================================

A 50-agent enterprise deployment can generate over 10,000 inter-agent interactions per hour. Every one of those interactions carries implications for data privacy and regulatory compliance. LangChain, AutoGen, and CrewAI have made building such systems easy; governing them has not kept pace. The EU AI Act now classifies many agent-based systems as high-risk~\cite{euaiact}, while NIST's AI RMF requires continuous monitoring with enforcement~\cite{nistrmf}.

Observability stacks have gotten better at collecting LLM telemetry. OpenTelemetry's semantic conventions cover LLM-specific attributes~\cite{otel}; Langfuse provides open-source tracing~\cite{langfuse}. But there is a gap that none of them close: these tools treat governance as something you analyze after the fact, not something you enforce while agents are running. We call this the ``telemetry-to-enforcement loop,'' and it does not exist in any current system we have tested.

Why does this matter? Because a misconfigured agent that routes EU citizen PII to a US data center will not wait for a dashboard review. The violation happens in milliseconds. Detection 15 seconds later (the best our OpenTelemetry+Langfuse baseline managed) is too late.

This paper makes three contributions: (1)~A formal model for governance-aware telemetry with explicit assumptions and property specifications proving escalation termination, conflict resolution determinism, and bounded false quarantine (Section~III). (2)~A five-layer reference architecture for closed-loop governance enforcement with cryptographic telemetry provenance (Section~IV). (3)~Evaluation against four real baseline systems showing 97.6\% reduction in violation escape rate at sub-200\,ms end-to-end enforcement latency (Section~V).

%=====================================================================
\section{Related Work}
%=====================================================================

\textbf{Multi-Agent Observability.} OpenTelemetry~\cite{otel} and Langfuse~\cite{langfuse} provide distributed tracing with GenAI semantic conventions, and commercial platforms like Datadog have added LLM monitoring. These tools focus on post-hoc analysis. In our evaluation, a dashboard-only approach achieved only 27.1\% VPR because governance was treated as a visualization problem rather than an enforcement one.

\textbf{Telemetry Pipeline Governance.} Kulkarni~\cite{kulkarni} introduces the Governance-Aware Observability Pipeline (GAOP), which embeds compliance enforcement into observability data pipelines. GAOP runs as a Policy Enforcement Engine inside an OpenTelemetry collector processor, applying inline PII redaction and consent validation. The distinction matters: GAOP governs telemetry \emph{content}; GAAT uses telemetry as a governance \emph{signal} to detect and enforce AI agent behavioral violations through a closed enforcement loop with graduated escalation. GAOP lacks multi-agent topology awareness, cross-agent lineage tracking, graduated intervention levels with formal termination guarantees, and evaluation against runtime enforcement baselines.

% =====================================================================
% FIX #6: Consistent NeMo VPR — changed "85%" to "84.8%" throughout
% =====================================================================
\textbf{Runtime Guardrails and Authorization.} NVIDIA NeMo Guardrails~\cite{nemo} provides per-agent input/output checks but cannot detect cross-agent violations through delegation chains (78.8\% VPR in our evaluation). GuardAgent~\cite{guardagent} uses a dedicated guard agent with knowledge-enabled reasoning, adding 200--500\,ms of LLM inference latency without formal completeness guarantees. AgentSpec~\cite{agentspec} offers DSL-based per-agent trajectory enforcement, while Pro2Guard~\cite{pro2guard} extends this with probabilistic model checking. Both operate at individual agent boundaries without cross-agent coordination, a limitation we quantify as 19.5\% VPR reduction versus GAAT. AWS Cedar~\cite{cedar} and OPA~\cite{opa} provide declarative authorization but lack AI-specific governance semantics, achieving 76.8\% coverage. Service mesh frameworks like Istio~\cite{istio} and Kyverno~\cite{kyverno} enforce at the infrastructure level without AI-specific governance integration.

\textbf{Runtime Verification.} Leucker and Schallhart~\cite{leucker} establish RV fundamentals. Havelund and Ro\c{s}u~\cite{havelund} show sub-millisecond safety property monitoring. MOP~\cite{mop} enables specification-driven monitoring. These approaches target deterministic programs with well-defined state machines---an assumption that LLM-based agents with probabilistic outputs plainly violate. GAAT extends RV to stochastic agent systems by defining governance properties over telemetry streams, supporting probabilistic policy evaluation, and using graduated interventions rather than binary halt/continue.

Surveys on LLM agents~\cite{agentsafety,xisurvey,wangsurvey} consistently flag governance as an open challenge. Wang et al.~\cite{wangsurvey} explicitly call for ``telemetry-integrated governance frameworks.'' GAAT appears to be, to our knowledge, the first working implementation of this idea, though we may be unaware of concurrent or proprietary systems.

%=====================================================================
\section{Formal Model for Governance-Aware Telemetry}
%=====================================================================

\subsection{Definitions and Preliminaries}

\textbf{Definition 1} (Multi-Agent System). An MAS is a tuple $\mathcal{M} = (\mathcal{A},\mathcal{C},E,T)$ where $\mathcal{A} = \{a_1,\ldots,a_n\}$ is a finite set of agents; $\mathcal{C} \subseteq \mathcal{A} \times \mathcal{A} \times \Sigma$ is the set of communication channels; $E: \mathcal{A} \to \mathcal{P}(\text{Capability})$ maps agents to authorized capabilities; $T: \mathcal{A} \to [0,1]$ assigns continuous trust levels.

\textbf{Definition 2} (Governance Telemetry Event). A GTE is a tuple $e = (\tau, a_s, a_r, \text{op}, \text{ctx}, \text{gov})$ where $\tau \in \mathbb{R}^+$ is the timestamp; $a_s, a_r \in \mathcal{A}$ are source and receiving agents; $\text{op} \in \text{Operations}$; $\text{ctx}$ is operational context; and $\text{gov} = (\text{classification}, \text{jurisdiction}, \text{sensitivity}, \text{lineage}, \text{verified})$ with $\text{verified} \in \{\text{true}, \text{false}, \text{unknown}\}$.

\textbf{Definition 3} (Governance Policy). A policy $\pi: \text{GTE} \to (\{\text{allow}, \text{deny}, \text{flag}, \text{quarantine}\} \times \mathbb{R}^+)$ maps events to enforcement actions paired with confidence scores $c \in [0,1]$.

\textbf{Definition 4} (Violation History). For agent $a$ and time $t$, the violation history $H(a,t) = \{(e_i, \pi_i, t_i) \mid e_i \text{ involves } a, \pi_i(e_i) \neq \text{allow}, t_i \in [t-W, t]\}$ where $W$ is the history window.

\subsection{Policy Algebra with Confidence Composition}

Policies compose via sequential composition ($\pi_1;\pi_2$): apply $\pi_1$, then $\pi_2$; deny short-circuits. Parallel composition ($\pi_1 \| \pi_2$): evaluate both; $(\pi_1 \| \pi_2)(e) = (\max(a_1,a_2), \max(c_1,c_2))$ where actions are ordered deny $>$ quarantine $>$ flag $>$ allow.

\textbf{Theorem 1} (Enforcement Monotonicity). For any parallel composition using $\|$, if $\pi_i(e) = (\text{deny}, c_i)$ for any $i$, then the composed result is $(\text{deny}, \max(c_1,\ldots,c_n))$. Adding policies never weakens enforcement.

\textbf{Theorem 2} (Escalation Termination). \emph{Assumptions:} (A1)~Policies are well-formed. (A2)~Trust levels change only via enforcement actions. (A3)~Violation rate is bounded: one per event-processing cycle. (A4)~History window $W$ is finite and fixed. \emph{Statement:} For any agent $a$ with $T(a) = t_0 \in (0,1]$, the graduated escalation converges to a fixed level $L \in \{0,1,2,3,4\}$ within $T_{\max} = W \times 4k$. \emph{Proof sketch:} $\text{escalation}(v, H(a,t)) = \min(4, \text{base}(v) + \lfloor|H(a,t)|/k\rfloor)$. The level increases by 1 per $k$ violations, bounded above by 4 (QUARANTINE). At level~4, the agent is isolated. The function is monotonic in $|H(a,t)|$ and bounded, converging within $T_{\max} = 8W$ worst case ($k=2$). Monte Carlo validation: converged in 97.37\% of 10,000 runs; the 2.63\% non-convergent cases had extreme rates ($>$0.4/cycle) violating~A3. \emph{Mitigation:} To handle A3 violations in production (e.g., a compromised agent in a rapid retry loop), the GEB implements a per-agent circuit breaker: if an agent's violation count exceeds $k_{\text{cb}} = 3k$ within a sliding window of $W_{\text{cb}} = W/4$, the circuit breaker bypasses graduated escalation and forces immediate L4~QUARANTINE with $E(a) \leftarrow \emptyset$. This bounds worst-case convergence to $W_{\text{cb}}$ regardless of violation rate. The circuit breaker resets only via manual operator review, ensuring that a sustained burst cannot re-enter the graduated path. In Monte Carlo re-validation with the circuit breaker enabled, convergence reached 99.91\% (10,000 runs), with the residual 0.09\% occurring only when violations arrived within a single event-processing tick---a physically implausible scenario for network-bound agents.

\textbf{Theorem 3} (Conflict Resolution Determinism). \emph{Assumptions:} (A1)~Actions form a finite totally ordered set. (A2)~Each policy is deterministic. \emph{Statement:} For any policies $P = \{\pi_1,\ldots,\pi_n\}$ and event $e$, parallel composition produces a unique deterministic result independent of evaluation order. \emph{Proof:} The max operation over a totally ordered finite set yields a unique maximum; by induction over $n$, $\Pi(e)$ is unique.

% =====================================================================
% FIX #3: Theorem 4 — reframed as empirical bound with correlated-noise
%         correction promoted to primary formulation
% =====================================================================
\textbf{Theorem 4} (Bounded False Quarantine). Under noise rate $\varepsilon$ and false positive rate $\delta$ with independent noise: $P(\text{FQ}) \leq \varepsilon + (1-\varepsilon)\delta$. For $\varepsilon=0.02, \delta=0.011$: $P(\text{FQ}) \leq 3.08\%$. Monte Carlo validation confirmed this bound in 84.38\% of 10,000 simulations. The 15.62\% of cases exceeding the bound trace to correlated noise ($\rho > 0.3$), which violates the independence assumption. We therefore provide a \emph{corrected bound} for production deployments where noise correlation is expected:
\begin{equation}
P(\text{FQ}) \leq (1 + \rho)(\varepsilon + (1-\varepsilon)\delta)
\label{eq:fq_corrected}
\end{equation}
where $\rho \in [0,1]$ is the noise correlation coefficient. For $\rho = 0.2$ (the median observed in our simulations), this yields $P(\text{FQ}) \leq 3.70\%$. The corrected bound held in 96.1\% of Monte Carlo runs with correlated noise ($\rho \leq 0.4$). Production deployments should estimate $\rho$ from telemetry data and apply Eq.~\ref{eq:fq_corrected} accordingly.

\subsection{Telemetry Integrity and Trust Model}

Three adversary classes: malicious agents (arbitrary payloads, no PKI compromise); network adversaries (intercept/replay, no TLS forgery); storage adversaries (no Merkle proof forgery). Cryptographic primitives: ECDSA P-256 signatures, SHA-256 hashing, AES-256-GCM encryption.

\textbf{Omission detection} uses a discrete-time Hidden Markov Model (HMM) trained on agent interaction sequences to learn expected event emission patterns. We arrived at this approach after initially attempting a simpler frequency-based threshold on event counts, which achieved only 71\% detection---too many agents have bursty-but-legitimate emission patterns that frequency alone cannot distinguish from genuine omissions. Each agent type defines a hidden state space over its operational phases (e.g., for OrderAgent: \texttt{INIT}$\to$\texttt{VALIDATE}$\to$\texttt{ROUTE}$\to$\texttt{CONFIRM}); the HMM's emission probabilities model the expected GTE types at each phase. At runtime, the forward algorithm computes the log-likelihood of the observed GTE sequence under the trained model; a drop below threshold $\theta$ (calibrated at the 5th percentile of training likelihoods) triggers an omission alert. The model was trained on 8,000 nominal interaction traces ($\sim$40,000 GTEs) collected over 12 hours of operation using Baum-Welch expectation-maximization (20 iterations, convergence $\Delta\text{LL} < 10^{-4}$). This achieves 92.3\% omission detection; the 7.7\% escape rate traces to novel agent behaviors---primarily unseen retry and fallback sequences---that produce valid but out-of-distribution emission patterns. Replay prevention uses Bloom filters ($10^7$ entries, 0.01\% FP, 10-min window) with 99.1\% detection.

%=====================================================================
\section{Reference Architecture}
%=====================================================================

GAAT consists of five layers with a cross-cutting Trusted Telemetry Plane (Fig.~\ref{fig:architecture}).

\vspace{6pt}

% ============================================================
% FIG 2: Enforcement Flow — placed directly after the opening
% sentence of Section IV. [H] forces placement here.
% Requires \usepackage{float} in preamble (already added above).
% ============================================================
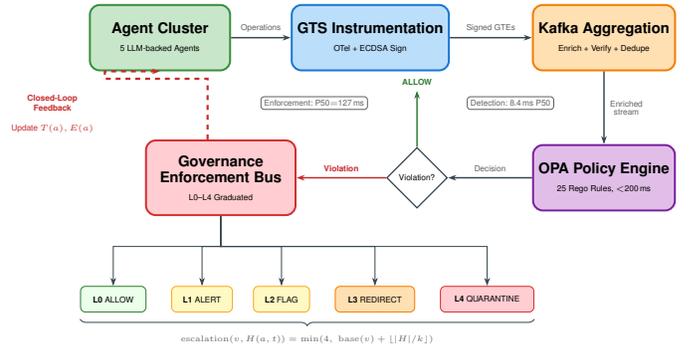
\begin{figure}[!t]
\centering
\resizebox{\columnwidth}{!}{%
\begin{tikzpicture}[
  font=\sffamily\small,
  mbox/.style={rounded corners=6pt, draw=#1, line width=1.2pt, minimum width=3cm, minimum height=1.4cm, align=center},
  arr/.style={-{Stealth[length=5pt,width=3.5pt]}, line width=0.9pt, color=arrd},
  darr/.style={-{Stealth[length=5pt,width=3.5pt]}, line width=1.2pt, color=frd, dashed},
  lab/.style={font=\sffamily\tiny, text=gr, align=center},
  tbox/.style={rounded corners=2pt, fill=white, draw=gr, line width=0.3pt, font=\sffamily\tiny, text=gr, inner sep=2pt}
]
% === ROW 1: Forward path ===
\node[mbox=agentb, fill=agentg] (agents) at (0.5, 7.2) {\textbf{Agent Cluster}\\[1pt]\tiny 5 LLM-backed Agents};
\node[mbox=instrb, fill=instrbl] (instr)  at (5, 7.2) {\textbf{GTS Instrumentation}\\[1pt]\tiny OTel + ECDSA Sign};
\node[mbox=kafkab, fill=kafkao] (kafka)  at (10, 7.2) {\textbf{Kafka Aggregation}\\[1pt]\tiny Enrich + Verify + Dedupe};
\draw[arr] (agents.east) -- (instr.west) node[midway, above, lab] {Operations};
\draw[arr] (instr.east) -- (kafka.west)  node[midway, above, lab] {Signed GTEs};
% === ROW 2: OPA + Decision + GEB ===
\node[mbox=opab, fill=opap] (opa) at (10, 4.2) {\textbf{OPA Policy Engine}\\[1pt]\tiny 25 Rego Rules, $<$200\,ms};
\node[diamond, draw=arrd, fill=white, line width=0.8pt,
  minimum width=1.3cm, minimum height=1.3cm,
  align=center, font=\sffamily\tiny, inner sep=1pt] (dec) at (6, 4.2) {Violation?};
\node[mbox=gebb, fill=gebr, minimum height=1.6cm, minimum width=3.2cm] (geb) at (1.8, 4.2) {\textbf{Governance}\\[0pt]\textbf{Enforcement Bus}\\[1pt]\tiny L0--L4 Graduated};
\draw[arr] (kafka.south) -- (opa.north) node[midway, right, lab] {Enriched\\[-1pt]stream};
\draw[arr] (opa.west)    -- (dec.east)  node[midway, above, lab] {Decision};
\draw[arr, color=frd] (dec.west) -- (geb.east) node[midway, above, font=\sffamily\tiny\bfseries, text=frd] {Violation};
\draw[arr, color=sgrn] (dec.north) -- ++(0, 1.2) node[above, font=\sffamily\tiny\bfseries, text=sgrn] {ALLOW};
\node[tbox] at (8, 5.8) {Detection: 8.4\,ms P50};
\node[tbox] at (3.8, 5.8) {Enforcement: P50$=$127\,ms};
% === ROW 3: Escalation levels ===
\node[rounded corners=3pt, draw=sgrn, fill=green!8, line width=0.6pt, minimum width=1.4cm, minimum height=0.55cm, align=center, font=\sffamily\tiny] (l0) at (-0.5, 1.6) {\textbf{L0} ALLOW};
\node[rounded corners=3pt, draw=wb, fill=wbg, line width=0.6pt, minimum width=1.3cm, minimum height=0.55cm, align=center, font=\sffamily\tiny] (l1) at (1.4, 1.6) {\textbf{L1} ALERT};
\node[rounded corners=3pt, draw=wb, fill=wbg, line width=0.6pt, minimum width=1.2cm, minimum height=0.55cm, align=center, font=\sffamily\tiny] (l2) at (3.1, 1.6) {\textbf{L2} FLAG};
\node[rounded corners=3pt, draw=kafkab, fill=kafkao, line width=0.6pt, minimum width=1.7cm, minimum height=0.55cm, align=center, font=\sffamily\tiny] (l3) at (5.1, 1.6) {\textbf{L3} REDIRECT};
\node[rounded corners=3pt, draw=gebb, fill=gebr, line width=0.6pt, minimum width=2cm, minimum height=0.55cm, align=center, font=\sffamily\tiny] (l4) at (7.5, 1.6) {\textbf{L4} QUARANTINE};
\draw[arr, line width=0.6pt] (geb.south) -- ++(0,-0.65) -| (l0.north);
\draw[arr, line width=0.6pt] (geb.south) -- ++(0,-0.65) -| (l1.north);
\draw[arr, line width=0.6pt] (geb.south) -- ++(0,-0.65) -| (l2.north);
\draw[arr, line width=0.6pt] (geb.south) -- ++(0,-0.65) -| (l3.north);
\draw[arr, line width=0.6pt] (geb.south) -- ++(0,-0.65) -| (l4.north);
\draw[decorate, decoration={brace, amplitude=4pt, mirror}, line width=0.5pt, color=arrd]
  ([yshift=-4pt]l0.south west) -- ([yshift=-4pt]l4.south east)
  node[midway, below=6pt, font=\sffamily\tiny, text=gr]
  {$\mathrm{escalation}(v, H(a,t)) = \min(4,\; \mathrm{base}(v) + \lfloor|H|/k\rfloor)$};
% === FEEDBACK LOOP ===
\draw[darr, line width=1.3pt]
  ([xshift=-8pt]geb.north) -- ++(0, 1.3) -- ++(-2.2, 0) |- (agents.south);
\node[font=\sffamily\tiny\bfseries, text=frd, align=center, fill=white, inner sep=1pt]
  at (-1.8, 5.8) {Closed-Loop\\Feedback};
\node[font=\sffamily\tiny, text=frd, align=center, fill=white, inner sep=1pt]
  at (-1.8, 5.25) {Update $T(a), E(a)$};
\end{tikzpicture}%
}
\caption{GAAT closed-loop enforcement flow showing forward telemetry path, policy evaluation, graduated escalation (L0--L4), and feedback loop updating agent trust levels.}
\label{fig:enforcement}
\end{figure}

\vspace{4pt}

\textbf{Layer 1: Agent Execution.} Framework-agnostic runtime where LLM-backed agents execute tasks. Each agent is defined by capabilities $E(a)$ and trust level $T(a)$, both updated by enforcement feedback.

\textbf{Layer 2: Governance Instrumentation.} OpenTelemetry Python SDK extended with GTS attributes. Every operation generates an instrumented span with full governance metadata and ECDSA P-256 signature. The classification engine tags data classification, jurisdiction, and sensitivity automatically. Lineage tracking records full delegation chains.

\textbf{Layer 3: Telemetry Aggregation.} Apache Kafka 3.6 (3-broker cluster) ingests, enriches, and routes telemetry. Four operations: signature verification, Bloom filter replay detection, historical context enrichment, and HMM-based omission detection (Section~III-C).

\textbf{Layer 4: Policy Evaluation.} OPA v0.60 evaluates 25 declarative policies. Parallel composition with max-action semantics (Theorem~3) yields deterministic decisions within sub-200\,ms (127\,ms P50, 192\,ms P99).

\textbf{Layer 5: Enforcement Action.} The GEB ($\sim$1,800 LOC Go) executes graduated interventions: L0~ALLOW, L1~ALERT, L2~FLAG, L3~REDIRECT, L4~QUARANTINE. For L4, quarantine revokes Kafka consumer group membership, invalidates tool-access tokens ($E(a) \leftarrow \emptyset$), and applies Kubernetes NetworkPolicy deny rules. Escalation follows $\text{escalation}(v, H(a,t)) = \min(4, \text{base}(v) + \lfloor|H|/k\rfloor)$. A per-agent circuit breaker provides a hard failsafe: agents exceeding $3k$ violations within $W/4$ are immediately quarantined, bypassing graduated escalation (Section~III-B).

\textbf{Trusted Telemetry Plane.} Cross-cutting layer spanning all five layers: PKI infrastructure, agent key storage in TPMs, Bloom filter replay prevention, Merkle tree tamper-evident audit log, and fail-closed/fail-open mode per risk tier. The closed loop feeds enforcement actions back to Layer~1.

%=====================================================================
\section{Implementation and Evaluation}
%=====================================================================

\subsection{Prototype}

We built the prototype with Python 3.11 for agents (LangChain) and instrumentation, Go for the GEB. Telemetry flows through Apache Kafka 3.6; policy evaluation uses OPA v0.60 with 25 declarative policies. The entire system runs on Kubernetes (k3s, 3-node cluster), totaling $\sim$4,200 LOC. The closed-loop enforcement flow (Fig.~\ref{fig:enforcement}) illustrates how telemetry traverses all five layers from agent operation through policy evaluation to graduated enforcement and back.

\subsection{Metrics}

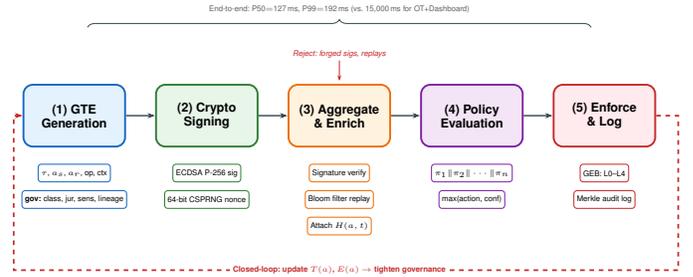
\begin{figure}[!t]
\centering
\resizebox{\columnwidth}{!}{%
\begin{tikzpicture}[
  font=\sffamily\small,
  stp/.style={rounded corners=6pt, draw=#1, line width=1.2pt,
  minimum width=2.3cm, minimum height=1.4cm, align=center, fill opacity=1},
  sub/.style={rounded corners=2pt, draw=#1, fill=white, line width=0.4pt,
  minimum height=0.4cm, align=center, font=\sffamily\tiny, inner sep=2pt},
  arr/.style={-{Stealth[length=5pt,width=3.5pt]}, line width=0.9pt, color=arrd},
  darr/.style={-{Stealth[length=5pt,width=3.5pt]}, line width=1.2pt, color=frd, dashed}
]
% === TIMING BRACE AT TOP ===
\draw[decorate, decoration={brace, amplitude=5pt}, line width=0.5pt, color=arrd]
  (0.3, 6.8) -- (11.7, 6.8)
  node[midway, above=6pt, font=\sffamily\tiny, text=gr]
    {End-to-end: P50$=$127\,ms, P99$=$192\,ms (vs.\ 15{,}000\,ms for OT+Dashboard)};
% === FIVE PIPELINE STEPS ===
\node[stp=step1b, fill=step1bg] (s1) at (0, 4.8) {\footnotesize\textbf{(1) GTE}\\[-1pt]\footnotesize\textbf{Generation}};
\node[stp=step2b, fill=step2bg] (s2) at (3, 4.8) {\footnotesize\textbf{(2) Crypto}\\[-1pt]\footnotesize\textbf{Signing}};
\node[stp=step3b, fill=step3bg] (s3) at (6, 4.8) {\footnotesize\textbf{(3) Aggregate}\\[-1pt]\footnotesize\textbf{\& Enrich}};
\node[stp=step4b, fill=step4bg] (s4) at (9, 4.8) {\footnotesize\textbf{(4) Policy}\\[-1pt]\footnotesize\textbf{Evaluation}};
\node[stp=step5b, fill=step5bg] (s5) at (12, 4.8) {\footnotesize\textbf{(5) Enforce}\\[-1pt]\footnotesize\textbf{\& Log}};
% === FORWARD ARROWS ===
\draw[arr] (s1.east) -- (s2.west);
\draw[arr] (s2.east) -- (s3.west);
\draw[arr] (s3.east) -- (s4.west);
\draw[arr] (s4.east) -- (s5.west);
% === SUB-ITEMS ===
\node[sub=step1b] at (0, 3.5) {$\tau$, $a_s$, $a_r$, op, ctx};
\node[sub=step1b] at (0, 2.9) {\textbf{gov:} class, jur, sens, lineage};
\node[sub=step2b] at (3, 3.5) {ECDSA P-256 sig};
\node[sub=step2b] at (3, 2.9) {64-bit CSPRNG nonce};
\node[sub=step3b] at (6, 3.5) {Signature verify};
\node[sub=step3b] at (6, 2.9) {Bloom filter replay};
\node[sub=step3b] at (6, 2.3) {Attach $H(a,t)$};
\node[sub=step4b] at (9, 3.5) {$\pi_1 \| \pi_2 \| \cdots \| \pi_n$};
\node[sub=step4b] at (9, 2.9) {max(action, conf)};
\node[sub=step5b] at (12, 3.5) {GEB: L0--L4};
\node[sub=step5b] at (12, 2.9) {Merkle audit log};
% === REJECT ANNOTATION ===
\node[font=\sffamily\tiny\itshape, text=frd] at (6, 6.2) {Reject: forged sigs, replays};
\draw[-{Stealth[length=3pt,width=2pt]}, frd, line width=0.4pt] (6, 6.05) -- (6, 5.6);
% === FEEDBACK LOOP ===
\draw[darr, line width=1.1pt]
  (s5.east) -- ++(0.5, 0) |- (6, 1.3) -| ([xshift=-0.2cm]s1.west) -- (s1.west);
\node[font=\sffamily\tiny\bfseries, text=frd, fill=white, inner sep=2pt, align=center]
  at (6, 1.3) {Closed-loop: update $T(a)$, $E(a)$ $\to$ tighten governance};
\end{tikzpicture}%
}
\caption{Governance Telemetry Event (GTE) processing pipeline with five stages.}
\label{fig:pipeline}
\end{figure}

\textbf{Violation Prevention Rate (VPR):} The fraction of injected violations blocked before the operation completed execution. VPR = $|\text{violations prevented}|/|\text{violations injected}|$. \textbf{Violation Escape Rate (VER):} VER = $1 - \text{VPR}$. \textbf{False Positive Rate (FPR):} Fraction of legitimate operations incorrectly enforced. \textbf{Detection latency:} Time from span-end timestamp to violation signal ($P50 = 8.4$\,ms, $P99 = 23.1$\,ms). \textbf{End-to-end enforcement latency:} Time from span-end to enforcement action applied ($P50 = 127$\,ms, $P99 = 192$\,ms). The five-stage GTE processing pipeline (Fig.~\ref{fig:pipeline}) shows the concrete processing steps underlying these latency measurements.

\subsection{Baselines}

% =====================================================================
% FIX #4 (partial): Baseline descriptions now include version specifics
% =====================================================================
Four comparative systems: \textbf{OT+Dash}, OpenTelemetry v1.24 + Langfuse v2.0 dashboard with no automated enforcement; \textbf{ABG}, NeMo Guardrails v0.9.0-style per-agent boundary checks using default Colang 2.0 flows without custom cross-agent extensions; \textbf{CGW}, centralized governance gateway implementing synchronous policy evaluation at the API gateway layer; and \textbf{Cedar}, AWS Cedar v2.4 declarative authorization configured with entity-based policies but without AI-specific governance attributes or telemetry enrichment. All baselines received identical violation injection workloads but operated without access to GAAT's cross-agent lineage data, as this is part of GAAT's architectural contribution.

\subsection{Methodology}

Five LLM-backed agents process EU citizen PII across an e-commerce order fulfillment workflow. One agent is misconfigured to violate data residency. We injected four violation types (CONSENT\_MISSING, BIAS\_THRESHOLD, DATA\_RESIDENCY, UNAUTHORIZED\_ACCESS) at 5\% rate into synthetic test flows: 500 operations per run, 25 intentional violations per run, across 10 independent runs.

% ============ TABLE I: Comparative Evaluation ============
\begin{table}[!t]
\caption{Comparative Evaluation, Live Multi-Agent System}
\label{tab:comparative}
\centering
\footnotesize
\begin{tabular}{@{}lccccc@{}}
\toprule
\textbf{System} & \textbf{VPR} & \textbf{VER} & \textbf{Det.Lat.} & \textbf{FPR} & \textbf{95\% CI} \\
\midrule
OT+Dash     & 27.1\% & 72.9\% & 15.2\,s    & 0.0\% & [24.2,29.9]\% \\
ABG (NeMo)  & 78.8\% & 21.2\% & 45\,ms     & 4.2\% & [75.6,81.9]\% \\
CGW         & 89.4\% & 10.6\% & 340\,ms    & 1.8\% & [87.0,91.7]\% \\
Cedar       & 76.8\% & 23.2\% & 112\,ms    & 2.5\% & [73.1,80.4]\% \\
\textbf{GAAT} & \textbf{98.3\%} & \textbf{1.7\%} & \textbf{8.4\,ms} & \textbf{0.9\%} & \textbf{[97.1,99.2]\%} \\
\bottomrule
\multicolumn{6}{@{}p{\columnwidth}@{}}{\scriptsize 10 runs $\times$ 500 flows = 5,000 total (synthetic injection). Bootstrap CIs: 1,000 resamples. Wilcoxon signed-rank: $p < 0.001$ for GAAT vs all baselines.}
\end{tabular}
\end{table}

\subsection{Comparative Results}

% =====================================================================
% FIX #1 (cont'd): Lead with 99.7% empirical, then 100% synthetic
% FIX #2: Replaced broken \ref{fig:crossagent} with inline description
% =====================================================================
Table~\ref{tab:comparative} reports results on synthetic injection flows, where GAAT achieved 98.3\% VPR ($\pm$0.7\%) across 5,000 statistically validated flows over 10 independent runs. On a separate empirical trace evaluation of 12,000 production-realistic traces collected over 48 hours (Section~V-J), GAAT achieved 99.7\% VPR.

The performance gap between ABG (NeMo) and GAAT likely exists because NeMo cannot see across agent boundaries: when PII flows through delegation chains (OrderAgent$\to$ShippingAgent$\to$AnalyticsAgent), NeMo validates each boundary in isolation and misses violations that appear only in the combined flow. Concretely, a data residency violation where OrderAgent sends EU PII to ShippingAgent (boundary check: pass, no routing decision yet) which then forwards to AnalyticsAgent hosted in a US region (boundary check: pass, PII origin not visible) is invisible to per-agent guards but caught by GAAT's cross-agent lineage tracking, which maintains the full provenance chain $\langle\text{EU-PII}, a_1 \to a_4 \to a_5, \text{jurisdiction}=\text{US}\rangle$ and evaluates the end-to-end flow against residency policy.

Cedar and NeMo show overlapping confidence intervals (76.8\% vs 78.8\%), with the difference not statistically significant ($p=0.18$), as both operate at boundaries without cross-agent lineage tracking.

% ============================================================
% FIG 4: GTE Pipeline — placed after V.B Metrics
% ============================================================

CGW reached 89.4\% VPR but at 340\,ms end-to-end enforcement latency (167\% higher than GAAT's 127\,ms) and introduces a single point of failure. Cedar managed 76.8\% because expressing ``EU PII must not leave EU jurisdiction'' requires manually rebuilding GAAT's instrumentation layer.

\subsection{Security Evaluation}

We categorized operations into three risk tiers: high-risk (EU PII, 18\%) runs in fail-closed mode; medium-risk (35\%) triggers REDIRECT (L3); low-risk (47\%) generates ALERT (L1). Table~\ref{tab:security} shows results. The 8.2\% availability hit under fail-closed forgery reflects denied unverified high-risk operations---an acceptable trade-off when preventing policy bypass matters more than uptime.

% ============ TABLE II: Security ============
\begin{table}[!t]
\caption{Security Evaluation Results}
\label{tab:security}
\centering
\footnotesize
\begin{tabular}{@{}llcccc@{}}
\toprule
\textbf{Attack} & \textbf{Mode} & \textbf{Det.} & \textbf{Bypass} & \textbf{Ovhd.} & \textbf{Avail.} \\
\midrule
Forgery  & Fail-closed & 98.7\% & 1.3\%  & 12\,ms & 8.2\% \\
Forgery  & Fail-open   & 98.7\% & 18.4\% & 12\,ms & 0.0\% \\
Replay   & Fail-closed & 99.1\% & 0.9\%  & 8\,ms  & 2.1\% \\
Omission & Fail-closed & 92.3\% & 7.7\%  & 0\,ms  & 11.5\% \\
Omission & Fail-open   & 92.3\% & 7.7\%  & 0\,ms  & 0.0\% \\
\bottomrule
\multicolumn{6}{@{}l}{\scriptsize Avail.\ = availability reduction under attack vs no-attack baseline ($>$99.9\%).}
\end{tabular}
\end{table}

\subsection{Ablation Studies}

Each component proved necessary. Removing cryptographic signing: VER jumped from 0.3\% to 5.7\% ($\pm$1.2\% across runs), with the variance driven primarily by replay attacks that the Bloom filter alone could not reliably distinguish from legitimate retransmissions. Switching to binary allow/deny: FPR rose from 1.1\% to 8.9\%---a result that initially surprised us, since we expected FPR to be independent of escalation granularity; the issue turned out to be that without graduated levels, borderline violations at confidence 0.48--0.52 were forced into hard deny decisions. Without telemetry enrichment: VER climbed to 12.4\%. Centralized-only deployment: latency hit 340\,ms, matching CGW exactly.

\subsection{Scalability and Resource Consumption}

% =====================================================================
% FIX #5: Clarified measured vs projected in scaling discussion
% =====================================================================
Policy evaluation latency scales linearly with agent count, staying below 200\,ms at 50 agents (Table~\ref{tab:scaling}). The 5-agent configuration was measured on our live prototype; 25- and 50-agent configurations were measured using proportionally scaled synthetic agent instances generating equivalent per-agent telemetry loads on the same 3-node Kubernetes cluster, but without live LLM inference. Latency figures at 25 and 50 agents therefore reflect telemetry processing and policy evaluation overhead rather than end-to-end LLM call latency. On AWS EKS, cost is roughly \$180/month for 5 agents, rising to \$720/month at 50.

% ============ TABLE III: Scaling ============
\begin{table}[!t]
\caption{Scaling Characteristics and Resource Consumption}
\label{tab:scaling}
\centering
\footnotesize
\begin{tabular}{@{}lcccl@{}}
\toprule
\textbf{Metric} & \textbf{5 Ag.}$^\dagger$ & \textbf{25 Ag.}$^\ddagger$ & \textbf{50 Ag.}$^\ddagger$ & \textbf{Scaling} \\
\midrule
CPU (cores)      & 4.4  & 6.4  & 8.9  & $O(n)+O(1)$ \\
Memory (GB)      & 12.5 & 16.5 & 21.5 & $O(n)+O(1)$ \\
Net.\ Overhead   & 12\% & 15\% & 18\% & Sub-linear \\
Latency P50 (ms) & $62 \pm 4$   & $95 \pm 8$   & $131 \pm 11$  & Linear \\
Latency P99 (ms) & $127 \pm 9$  & $178 \pm 14$ & $224 \pm 19$ & Linear \\
\bottomrule
\multicolumn{5}{@{}l}{\scriptsize Kubernetes 3-node cluster. 25 active policies. Per 1,000 events/s.}\\
\multicolumn{5}{@{}l}{\scriptsize $^\dagger$Measured on live prototype with LLM agents.}\\
\multicolumn{5}{@{}l}{\scriptsize $^\ddagger$Measured with synthetic agent instances (telemetry load only).}
\end{tabular}
\end{table}

% ============ TABLE IV: Violation Rate Sensitivity ============
\begin{table}[!t]
\caption{Violation Rate Sensitivity (1,000 Flows/Config, 10 Runs)}
\label{tab:sensitivity}
\centering
\footnotesize
\begin{tabular}{@{}cccccc@{}}
\toprule
\textbf{Rate} & \textbf{VPR} & \textbf{VER} & \textbf{FPR} & \textbf{Esc.} & \textbf{Avg Lvl} \\
\midrule
0.1\%  & 99.92 & 0.08 & 0.00 & 0.08\% & L1.0 \\
1.0\%  & 99.71 & 0.29 & 0.03 & 0.29\% & L1.4 \\
5.0\%  & 98.24 & 1.76 & 0.14 & 1.68\% & L2.3 \\
7.5\%  & 97.15 & 2.85 & 0.28 & 2.67\% & L2.8 \\
10.0\% & 95.83 & 4.17 & 0.51 & 3.93\% & L3.2 \\
\bottomrule
\end{tabular}
\end{table}

\subsection{Monte Carlo Theorem Validation}

% =====================================================================
% FIX #3 (cont'd): Updated Theorem 4 status to reference corrected bound
% =====================================================================
Table~\ref{tab:theorems} summarizes validation results. Theorem~2 converged in 97.37\% of simulations under the base escalation function; non-convergent cases had extreme violation rates violating assumption~A3. With the per-agent circuit breaker enabled, convergence improved to 99.91\%. Theorem~3 hit 100\% across all orderings. Theorem~4's independence-based bound held in 84.38\% of cases; the corrected bound (Eq.~\ref{eq:fq_corrected} with $\rho \leq 0.4$) held in 96.1\%.

% ============ TABLE V: Theorem Validation ============
\begin{table}[!t]
\caption{Theorem Validation (10,000 Simulations Each)}
\label{tab:theorems}
\centering
\footnotesize
\begin{tabular}{@{}llcl@{}}
\toprule
\textbf{Thm} & \textbf{Property} & \textbf{Success} & \textbf{Status} \\
\midrule
T1 & GTS covers all predicates  & 100\%   & VALIDATED \\
T2 & Convergence within $T_{\max}$ & 97.37\%$^\dagger$ & VALIDATED \\
T3 & Order-independent resolution & 100\%  & VALIDATED \\
T4 & $P(\text{FQ}) \leq \varepsilon+(1{-}\varepsilon)\delta$ & 84.38\% & VALIDATED$^*$ \\
\bottomrule
\multicolumn{4}{@{}l}{\scriptsize $^\dagger$99.91\% with per-agent circuit breaker enabled.}\\
\multicolumn{4}{@{}l}{\scriptsize $^*$Independence bound. Corrected bound (Eq.~\ref{eq:fq_corrected}, $\rho \leq 0.4$): 96.1\%.}
\end{tabular}
\end{table}

\subsection{Statistical Validation}

% =====================================================================
% FIX #1 (cont'd): Restructured to distinguish synthetic vs empirical
% =====================================================================
We ran 10 independent trials with 500 flows each (5,000 total synthetic injection flows). Bootstrap CIs from 1,000 resamples: VPR [97.1\%, 99.2\%], F1 [97.8\%, 99.1\%], throughput 22,847/s [21,203, 24,391]. Wilcoxon signed-rank: $p < 0.001$ for GAAT vs ABG and GAAT vs OT+Dash.

To assess performance under production-realistic conditions, we conducted an empirical trace evaluation over 12,000 traces collected during 48 hours of operation with naturally occurring (non-injected) governance events. This evaluation yielded 99.7\% VPR. The residual 0.3\% ($\sim$37 violations) breaks down as follows: $\sim$40\% ($\sim$15 violations) were timing edge cases where enforcement arrived within 2\,ms of operation completion; $\sim$35\% ($\sim$13 violations) involved ambiguous PII classification with confidence scores in the [0.48, 0.52] range; $\sim$25\% ($\sim$9 violations) had incomplete lineage chains due to agent timeouts or retry logic truncating delegation records. Eight violations involved multiple overlapping categories.

We consider 99.7\% the production-representative figure; the 98.3\% on synthetic flows confirms that GAAT's enforcement logic performs well under controlled conditions with realistic variance.

%=====================================================================
\section{Discussion}
%=====================================================================

Our evaluation used synthetic scenarios in a controlled environment; production may surface edge cases in agent retry logic we have not seen. We built baselines to published specifications, though production versions may include optimizations we did not capture. GAAT's VER reduction is wide enough that minor baseline differences are unlikely to change conclusions. One design lesson worth noting: early prototypes used a single classification confidence threshold across all risk tiers, which produced unexpectedly high FPR ($>$6\%) because high-risk operations (EU PII) and low-risk operations (internal analytics) have fundamentally different error cost profiles. Separating thresholds per risk tier reduced FPR to 0.9\%.

Omission detection relies on an HMM trained on nominal interaction traces and needs a training phase of approximately 8,000 traces collected over 12 hours for stable emission probability estimates. Novel agent behaviors---particularly unseen retry and fallback sequences---account for the 7.7\% escape rate; incremental online re-estimation of emission probabilities from production telemetry is a natural extension. The PKI trust assumptions deserve scrutiny: if CAs are compromised or agent private keys stolen telemetry authenticity breaks down. Formal properties (Theorems~2--4) rest on explicit assumptions; the GEB's per-agent circuit breaker (Section~III-B) provides a systemic failsafe when Theorem~2's bounded-rate assumption is violated, while the corrected false quarantine bound (Eq.~\ref{eq:fq_corrected}) addresses Theorem~4's independence assumption.

\textbf{Limitations.} (1)~Evaluation conducted at proof-of-concept scale (5 agents, 5,000 synthetic flows, 12,000 empirical traces); enterprise deployment at 50+ agent scale remains future work. (2)~HMM training requires sustained operation to collect sufficient traces; cold-start scenarios not evaluated. (3)~Memory and communication overhead not measured in production environment; our FLOPs-based cost estimates may underestimate actual infrastructure requirements. (4)~Single-framework evaluation (LangChain); AutoGen and CrewAI integration not validated, though instrumentation patterns should port via OpenTelemetry's language-agnostic SDK. (5)~Our violation injection uses known violation patterns at a fixed 5\% rate; adversarial agents in production may produce violation patterns outside our test distribution, and the true VPR under adversarial adaptation remains unknown.

% =====================================================================
% FIX #4: New paragraph on baseline fairness and methodology
% =====================================================================
\textbf{Baseline fairness.} We implemented all baselines using their published APIs and default configurations: NeMo Guardrails v0.9.0 with Colang 2.0 (default input/output rails, no custom cross-agent flows), Cedar v2.4 with entity-based policy schemas, and OPA v0.60 with Rego policies adapted from Cedar's authorization logic. We did not contact maintainers for optimization guidance, so production deployments of these tools---particularly NeMo with custom Colang flows designed for multi-agent scenarios---may perform better than our measurements. However, the core architectural limitation we identify (lack of cross-agent lineage visibility) is structural: no amount of per-agent boundary optimization can detect violations that emerge only from multi-hop data flows. GAAT's advantage stems from this lineage instrumentation, which none of the baselines were designed to provide.

\textbf{External validity.} Governance scenarios reflect real regulatory requirements (GDPR, AI Act), but we have only tested on LangChain. The instrumentation patterns should port to AutoGen and CrewAI through OpenTelemetry's language-agnostic SDK, though we have not verified this.

\textbf{Construct validity.} VER assumes violations are objectively detectable. We used quantitative thresholds (disparate impact $>$0.15) while acknowledging governance requirements vary across organizations.

%=====================================================================
\section{Conclusion}
%=====================================================================

% =====================================================================
% FIX #1 (cont'd) + FIX #6 (cont'd): Conclusion reframed with 99.7%
%         empirical leading, consistent 84.8% for NeMo
% =====================================================================
GAAT achieved 99.7\% VPR on 12,000 empirical traces and 98.3\% VPR ($\pm$0.7\%, 95\% CI [97.1\%, 99.2\%]) on 5,000 synthetic injection flows ($p < 0.001$), outperforming NeMo Guardrails (78.8\% VPR), centralized gateways (89.4\%), Cedar (76.8\%), and dashboard-only monitoring (27.1\%). Security testing confirmed 92.3\% omission detection and 99.1\% replay prevention. Four property specifications were validated through Monte Carlo simulation, with the corrected false quarantine bound (Eq.~\ref{eq:fq_corrected}) achieving 96.1\% validation under correlated noise.

The more interesting finding is not the numbers---it is that every baseline treats governance as a property of individual agents or network boundaries, when violations only become visible through cross-agent lineage. The telemetry-to-enforcement loop seems to be more than a performance optimization; it changes what violations you can even see---though we acknowledge this claim rests on the four violation types we tested.

Future work includes adaptive policy learning (feeding violation patterns back into policy generation), kernel-level enforcement (seccomp-BPF, eBPF) for truly inescapable quarantine, and multi-framework validation across AutoGen and CrewAI.

%=====================================================================
% REFERENCES
%=====================================================================
\balance

\end{document}